\documentclass[letter]{aa}     

\usepackage{graphicx}
\usepackage{txfonts}
\usepackage{lipsum}
\usepackage{subcaption}         
\usepackage{lscape}             
\usepackage{placeins}           

\usepackage{microtype}

\usepackage{ulem}

\usepackage[colorlinks = true,linkcolor = blue,urlcolor  = blue,citecolor = blue,anchorcolor = blue]{hyperref}

\begin{document}

   \title{Probing the inner disk edge of rotating neutron stars with dipole excitations in finite nuclei}
   

%
%
%

   \author{P.S. Koliogiannis\corrauth{pkoliogi@phy.hr}        
        \and N. Paar\email{npaar@phy.hr}
        }

   \institute{Department of Physics, Faculty of Science, University of Zagreb, Bijeni\v cka cesta 32, 10000, Zagreb, Croatia.}

   \date{\today}


\abstract
   {The nuclear equation of state shapes both the structure of rapidly rotating neutron stars and the orbital motion in their surrounding spacetime. Its uncertainty affects the mass and radius at which the innermost stable circular orbit reaches the stellar surface, a condition that defines the transition between disks whose inner edge is set by the surface and those truncated by orbital instability.}
   {We quantify how measurements of the electric dipole polarizability of neutron-rich nuclei reduce the uncertainty in this transition and in the orbital properties at the disk inner edge.}
   {We compute uniformly rotating neutron stars in general relativity for a family of equations of state that systematically vary the symmetry energy, derived from relativistic energy density functionals with density-dependent point-coupling interactions. We compare the results for the full family with those for the subset consistent with the measured dipole polarizabilities of finite nuclei.}
   {The dipole polarizability constraint reduces the spread in transition mass and equatorial radius by factors of about seven to ten across the spin frequencies considered. Along this transition, it also confines the orbital frequency at the disk inner edge to a band narrower than $20$~Hz at each spin.}
   {Combined with a measured neutron-star spin, a kilohertz quasi-periodic oscillation interpreted as disk orbital motion further restricts the mass--radius region, allowing an independent mass estimate to probe the symmetry-energy constraint. The dipole response of finite nuclei thus informs both the structure of rotating neutron stars and the orbital properties of their inner accretion disks.}


   \maketitle

\nolinenumbers

\section{Introduction}
Rapidly rotating neutron stars are among the few environments in which the properties of matter at supranuclear densities leave an imprint on orbital dynamics in strong gravitational fields~\citep{Klis_2000,Psaltis-2008}. Their structure depends on the equation of state (EOS;~\citealt{Oertel2017}), while rotation modifies both hydrostatic equilibrium, through centrifugal support, and the surrounding spacetime~\citep{Paschalidis_2017}. Together, the EOS and rotation govern the equatorial radius and the location of the prograde innermost stable circular orbit (ISCO) relative to the stellar surface. For a thin disk with negligible magnetic truncation, the inner edge is set by the stellar surface when stable circular orbits reach it, and by an external ISCO otherwise. This edge sets the highest Keplerian frequency available in the disk and the orbital binding energy that matter can release before reaching it~\citep{Thampan_1998,Sibgatullin_1998}. Kilohertz quasi-periodic oscillations (kHz QPOs) in low-mass X-ray binaries may trace orbital motion near this edge and, under that interpretation, constrain the stellar mass and radius and hence the nuclear EOS~\citep{Miller_1998}.

Early work by~\citet{Kluzniak-1985} examined the relation between the ISCO and the stellar surface. Subsequent general-relativistic studies established its sensitivity to the EOS and rotation, including the role of the spin-induced quadrupole moment~\citep{Cook-1994,PhysRevD.58.104011,Thampan-1999,Berti-2004}. The interpretation of the highest kHz QPO frequencies as orbital motion near the ISCO motivated early constraints on neutron-star masses and radii~\citep{Kaaret_1997,Zhang_1997,Kluzniak_1998,Miller_1998,Schaab_1999}. More recent studies have examined the appearance of an external ISCO and its orbital properties~\citep{Torok-2014,PhysRevD.96.024046}, developed approximately EOS-insensitive ISCO relations~\citep{Pappas-2015,Luk_2018}, and compared the highest observed QPO frequencies with the maximum orbital frequencies predicted for rotating neutron stars~\citep{van_Doesburgh-2018}.

Despite this progress, predictions of the mass and spin at which the ISCO reaches the stellar surface, hereafter the surface--ISCO transition, remain limited by uncertainty in the EOS, which affects both the stellar radius and the exterior spacetime. A major source of this uncertainty is the density dependence of the symmetry energy, which strongly influences neutron-star radii~\citep{Lattimer_2001}. Measurements of the electric dipole polarizability of neutron-rich nuclei constrain this dependence near and below nuclear saturation density~\citep{ROCAMAZA201896}, while astrophysical observations provide complementary information at the higher densities reached inside neutron stars~\citep{Yunes_2022,ymsq-cfcw}. These nuclear measurements have already narrowed predictions for nonrotating stars~\citep{Koliogiannis_2026,9mnb-t3yz}.

In this Letter, we introduce constraints on the surface--ISCO transition of rapidly rotating neutron stars and the orbital properties at the disk inner edge, by utilising measurements of the electric dipole polarizability of neutron-rich nuclei. We model uniformly rotating neutron stars in general relativity, adopting EOSs derived from a family of relativistic energy density functionals with density-dependent point-coupling (DD--PC) interaction. These functionals are calibrated to the same nuclear ground-state data, with the symmetry energy at saturation density systematically varied across the family~\citep{Yuksel2021,Koliogiannis_2026}. By comparing the full family with the subset consistent with the measured polarizabilities, we quantify the reduction in the predicted spread of transition masses, radii, and orbital properties. Combined with a measured stellar spin and a kHz QPO interpreted as orbital motion near the disk inner edge, these predictions restrict the allowed mass--radius region of accreting neutron stars, linking a nuclear observable to the dynamics of the inner accretion disk.

\section{Methodology}
To quantify how constraints from nuclear collective excitations refine the surface--ISCO transition, we combine fully relativistic models of uniformly rotating neutron stars with the DD--PC-J29--J36 family of EOSs~\citep{Yuksel2021}. These EOSs are derived from relativistic energy density functionals with DD--PC interactions~\citep{NIKSIC20141808}. Their construction and applications to neutron-star properties are described in~\citet{KOLIOGIANNIS2025139362,Koliogiannis_2026,9mnb-t3yz}. Across the family, the symmetry energy at saturation density spans $J=29$--$36$~MeV, with corresponding slope parameters $L=29$--$94.1$~MeV. For nonrotating $1.4~M_\odot$ stars, the predicted radii span $R_{1.4}=11.81$--$12.90$~km, while all EOSs support a maximum nonrotating mass of at least $2~M_\odot$.

Using measurements of the electric dipole polarizability in $^{48}$Ca~\citep{PhysRevLett.118.252501}, $^{68}$Ni~\citep{PhysRevLett.111.242503}, $^{120}$Sn~\citep{PhysRevC.92.031305}, and $^{208}$Pb~\citep{PhysRevLett.107.062502}, collectively denoted as the CNSP-4 set, \citet{Koliogiannis_2026} inferred $J=31.34\pm0.79$~MeV. The resulting predictions were found to be compatible with the astrophysical constraints considered in that work, including pulsar mass measurements~\citep{doi:10.1126/science.1233232,Fonseca_2021,Romani_2022}, mass--radius inferences from NICER~\citep{Miller_2019,Riley_2019,Miller_2021,Riley_2021,Choudhury_2024,Salmi_2024}, and tidal deformability constraints from GW170817~\citep{PhysRevLett.119.161101,PhysRevX.9.011001}.

For each EOS, we compute uniformly rotating configurations in general relativity with the \texttt{RNS} code~\citep{Stergioulas-1995}. We retain only those between the mass-shedding limit and the turning points of the mass along sequences of constant angular momentum, which we adopt as the onset of secular axisymmetric instability~\citep{Friedman-1988}.
Along sequences of constant spin frequency $f$, we use the exterior spacetime of each configuration to locate the prograde ISCO relative to the stellar surface. The inner edge of a thin disk then lies at the circumferential radius
\begin{align}
    \text{stellar surface:}& \quad R_{\rm in}=R_{\rm eq}, \nonumber \\
    \text{external ISCO:}& \quad R_{\rm in}=R_{\rm ISCO}>R_{\rm eq},
\end{align}
where $R_{\rm eq}$ and $R_{\rm ISCO}$ are the circumferential radii of the stellar equator and the prograde ISCO, respectively. The condition $R_{\rm ISCO}=R_{\rm eq}$ defines the transition gravitational mass $M_{\rm tr}(f)$. At $R_{\rm in}$, we evaluate the prograde orbital frequency $\nu_{\rm in}=\Omega_{\rm orb}(R_{\rm in})/(2\pi)$ and the orbital binding efficiency $\eta_{\rm in}=1-E_{\rm in}$, where $E_{\rm in}$ is the conserved specific orbital energy normalized to the rest-mass energy. In the nonrotating limit, where the exterior spacetime is Schwarzschild, these reduce to
\begin{equation}
    \nu_{\rm in}=\frac{1}{2\pi}\sqrt{\frac{GM}{R_{\rm in}^3}}, \qquad
    E_{\rm in}=\frac{1-2GM/(R_{\rm in}c^2)}{\sqrt{1-3GM/(R_{\rm in}c^2)}},
\end{equation}
with $R_{\rm ISCO}=6GM/c^2$~\citep{Shapiro-1983}, so the transition occurs at a stellar compactness of $GM_{\rm tr}/(R_{\rm eq}c^2)=1/6$. Comparing the full family with the CNSP-4 band for the symmetry energy then isolates the effect of the nuclear constraint on $M_{\rm tr}(f)$ and the associated orbital properties.

\begin{figure}
    \centering
    \includegraphics[width=0.95\columnwidth]{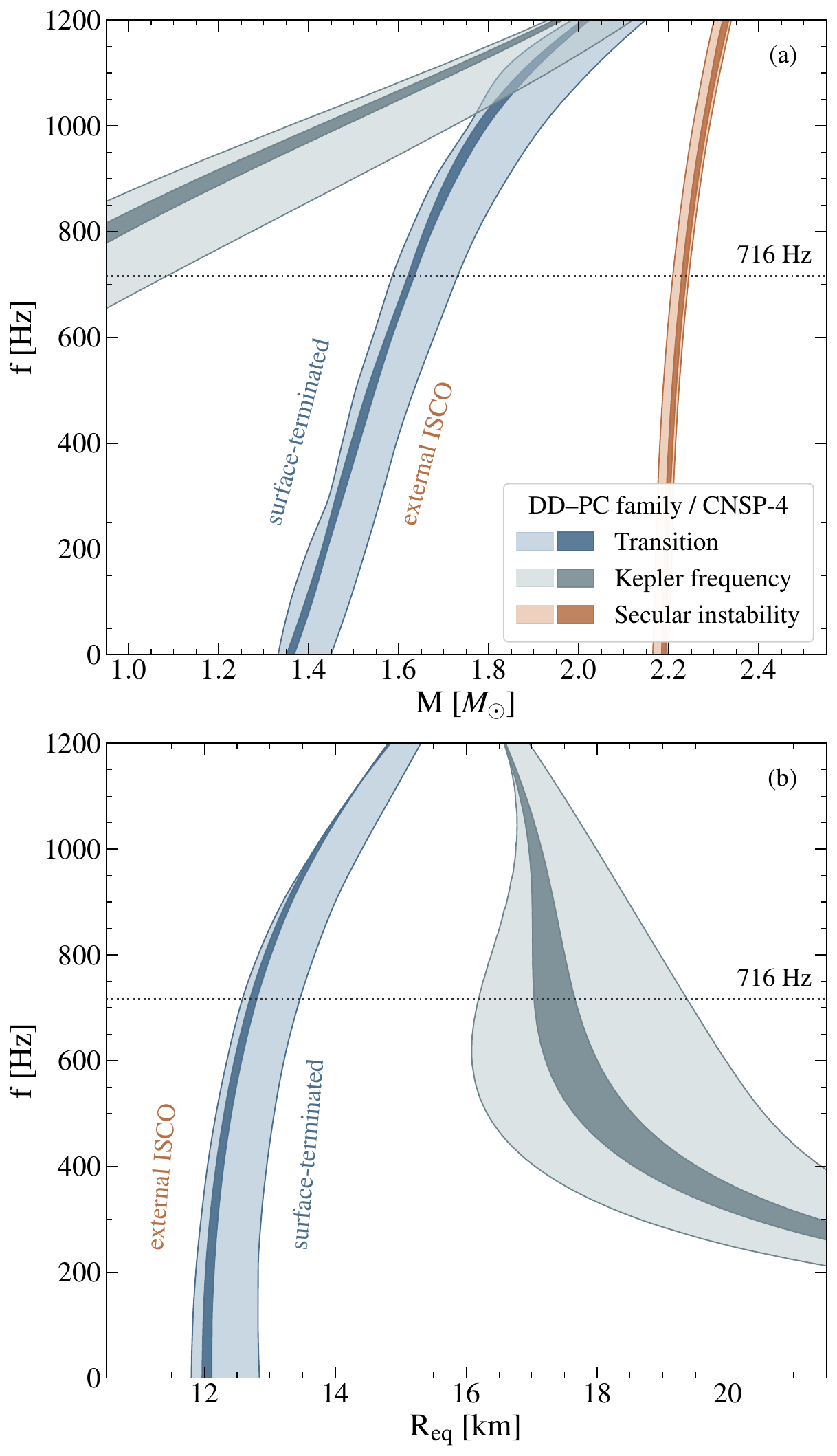}
    \caption{Stellar spin frequency $f$ as a function of (a) gravitational mass $M$ and (b) circumferential equatorial radius $R_{\rm eq}$ for the DD--PC EOSs. The bands show the Kepler (mass-shedding) limit, the surface--ISCO transition, and the adopted secular axisymmetric-instability boundary. At fixed spin, they appear in this order from left to right in panel (a); in panel (b), the transition lies at smaller radii than the Kepler limit. Lighter shading spans the full EOS family, while darker shading denotes the CNSP-4-constrained range. The dotted horizontal line marks $f=716$~Hz.}
    \label{fig:freq_mass_radius}
\end{figure}

\section{Results and discussion}
Figure~\ref{fig:freq_mass_radius} displays the gravitational mass $M$ and equatorial radius $R_{\rm eq}$ at the surface--ISCO transition across the stellar spin frequencies considered, together with the mass-shedding limit and the adopted stability boundary. Depending on spin, the full DD--PC family yields spreads of approximately $0.12$--$0.2~M_\odot$ in transition mass and $0.5$--$1$~km in transition radius. Restricting the family to the CNSP-4 band reduces these spreads to below $0.025~M_\odot$ and $0.15$~km, respectively, corresponding to reductions of approximately $85$--$90\%$ (factors of about seven to ten) across the spins considered. This refinement narrows the range of masses and radii over which the choice of EOS determines whether the disk reaches the stellar surface or is limited by an external ISCO. At fixed mass, it also narrows the spin interval over which the disk regime remains EOS dependent.

For illustration, we adopt $f=716$~Hz, the spin frequency of PSR~J1748$-$2446ad, the fastest-spinning neutron star known~\citep{doi:10.1126/science.1123430}. At this spin, the transition occurs at $M\simeq1.586$--$1.731~M_\odot$ and $R_{\rm eq}\simeq12.58$--$13.46$~km across the full DD--PC family. Within the CNSP-4 band, these intervals contract to $M\simeq1.620$--$1.635~M_\odot$ and $R_{\rm eq}\simeq12.70$--$12.80$~km, reducing the mass and radius spreads by about $90\%$ and $88\%$, respectively.

\begin{figure}
    \centering
    \includegraphics[width=0.95\columnwidth]{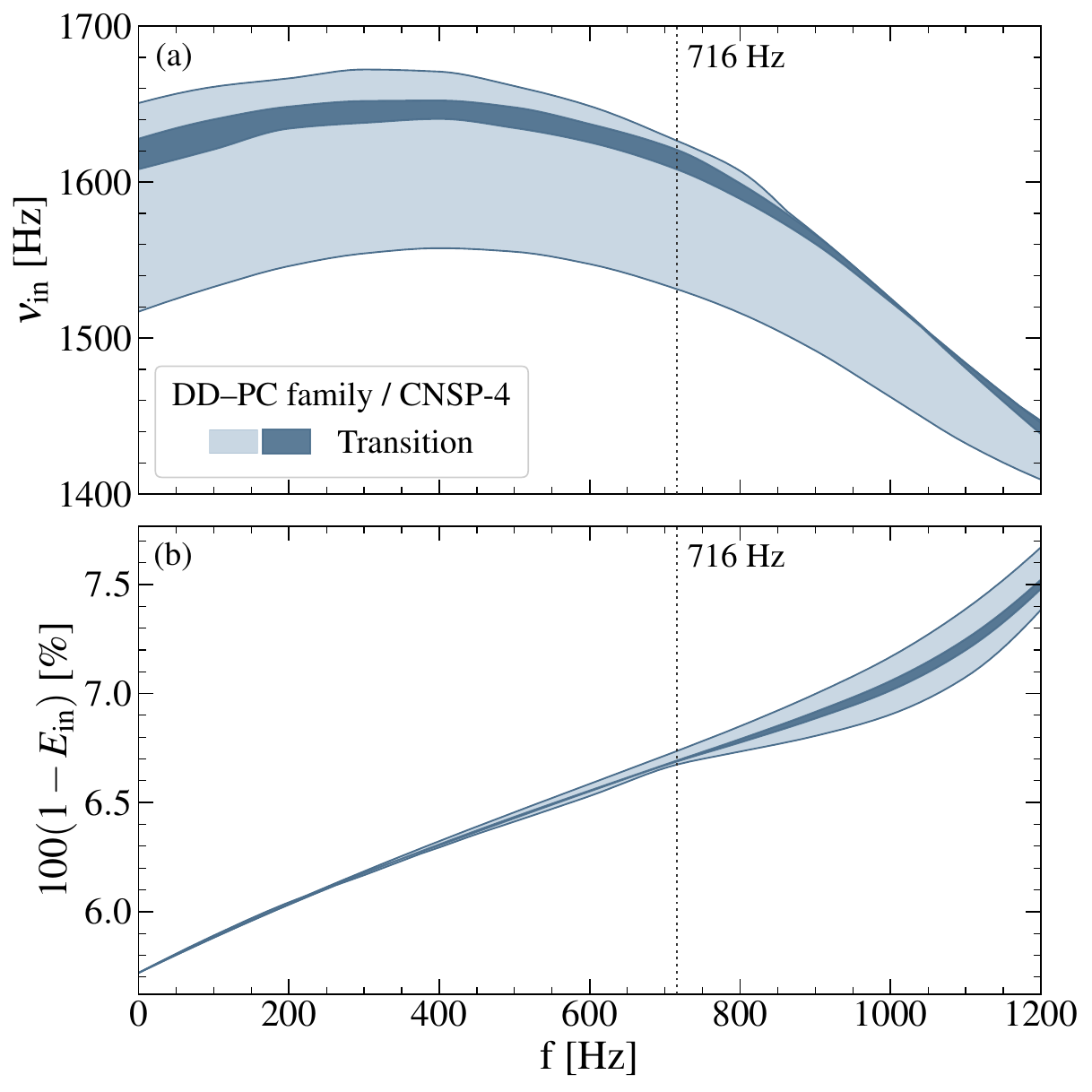}
    \caption{(a) Inner-edge orbital frequency $\nu_{\rm in}$ and (b) binding efficiency $\eta_{\rm in}$ as functions of stellar spin frequency $f$ along the surface--ISCO transition for the DD--PC EOSs. Lighter shading spans the full EOS family, while darker shading denotes the CNSP-4-constrained range. The dotted vertical line marks $f=716$~Hz.}
    \label{fig:orbital_properties}
\end{figure}

Figure~\ref{fig:orbital_properties} traces the inner-edge orbital frequency $\nu_{\rm in}$ and binding efficiency $\eta_{\rm in}$ along the surface--ISCO transition. The frequency sets the highest Keplerian frequency available in the assumed disk, while the efficiency measures the fraction of rest-mass energy available for release before matter reaches its inner edge. At each spin considered, CNSP-4 constraint confines $\nu_{\rm in}$ to a band narrower than $20$~Hz, compared with widths of approximately $38$--$134$~Hz for the full DD--PC family, corresponding to an average reduction of about $89\%$. Along the transition, $\nu_{\rm in}$ increases with spin, reaches a maximum at $f\approx300$--$400$~Hz, and decreases at faster rotation. The initial rise reflects the influence of frame dragging, which tends to reduce the prograde ISCO radius at fixed mass, shifting the surface--ISCO transition toward larger masses while $R_{\rm eq}$ remains nearly constant (Fig.~\ref{fig:freq_mass_radius}). At higher spin, centrifugal expansion and the outward contribution of the spin-induced quadrupole become increasingly important~\citep{PhysRevD.58.104011}. Along the transition, the equatorial radius then grows more rapidly, and $\nu_{\rm in}$ declines despite the continued increase in mass.

The binding efficiency rises with spin from the EOS-independent nonrotating value of $1-\sqrt{8/9}\approx5.72\%$. Its EOS-dependent spread grows with spin, reaching about $0.3$ percentage points for the full family, whereas within CNSP-4 band it remains at most about $0.05$ percentage points. At $716$~Hz, the spread in $\nu_{\rm in}$ contracts from approximately $95$ to $13$~Hz, and that in $\eta_{\rm in}$ from approximately $0.062$ to $0.004$ percentage points. For a star at the transition, a kHz QPO attributed to orbital motion at the disk inner edge should therefore fall within a predicted window narrower than $20$~Hz. Relating the binding efficiency to an observed luminosity additionally requires the accretion rate and a model for emission from the stellar surface or boundary layer~\citep{Thampan_1998}.

\begin{figure}
    \centering
    \includegraphics[width=0.95\columnwidth]{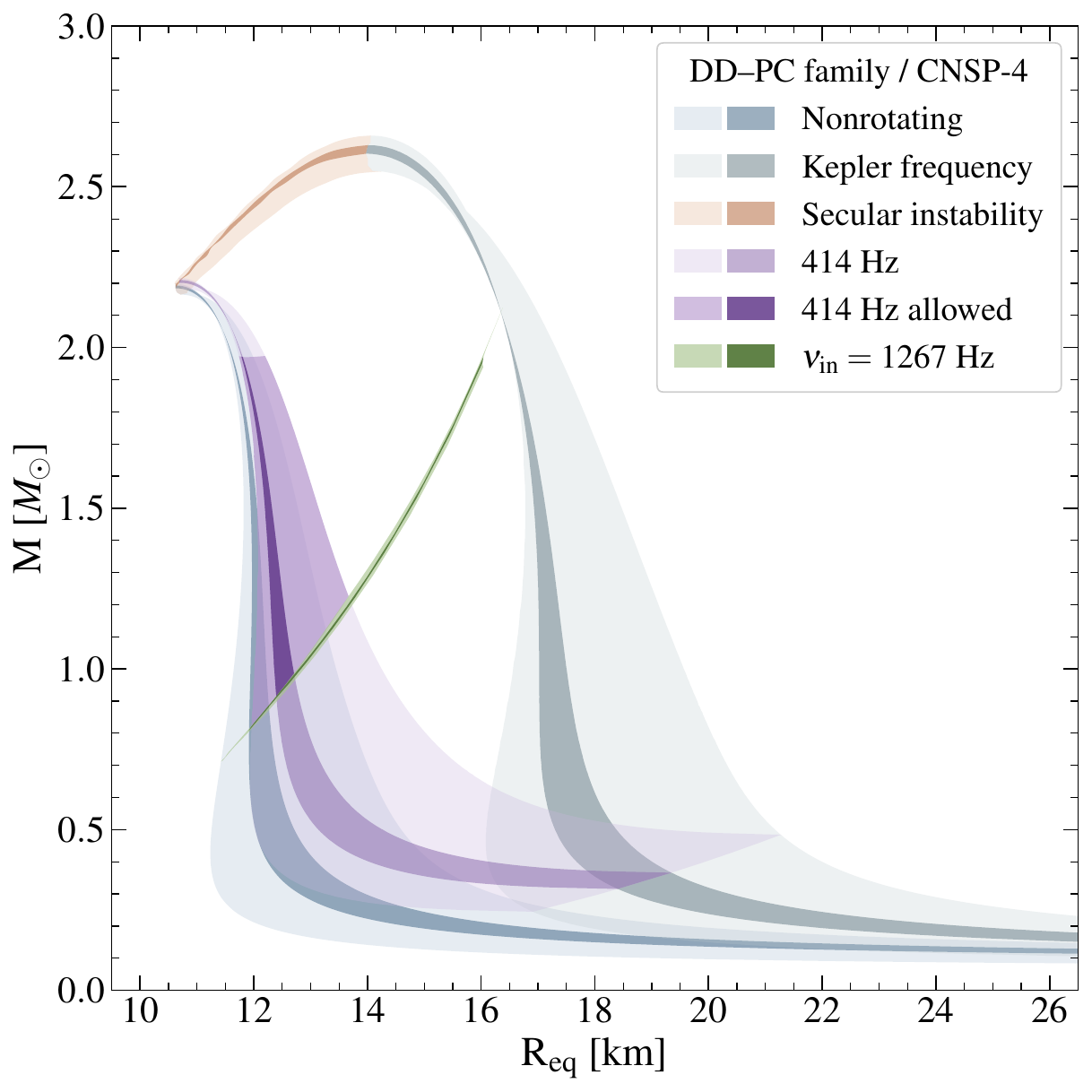}
    \caption{Gravitational mass $M$ as a function of circumferential equatorial radius $R_{\rm eq}$ for the DD--PC EOSs. The nonrotating and Kepler (mass-shedding) sequences, the adopted secular axisymmetric-instability boundary, and the $414$~Hz sequences are shown. Highlighted portions of the $414$~Hz band satisfy $\nu_{\rm in}\geq1267$~Hz, the orbital-frequency condition adopted for 4U~0614+09. The $\nu_{\rm in}=1267$~Hz band marks this threshold across the spin frequencies considered. Lighter shading spans the full EOS family, while darker shading denotes the CNSP-4-constrained range.}
    \label{fig:mass_radius}
\end{figure}

Figure~\ref{fig:mass_radius} shows the mass--radius regions occupied by the DD--PC family from the nonrotating limit to mass shedding. For nonrotating stars, CNSP-4 constraint from dipole polarizability narrows the mass--radius relation, as established by~\citet{Koliogiannis_2026}. Increasing the spin extends the configurations to larger equatorial radii at a given mass, while the CNSP-4 band remains substantially narrower than that of the full family, including along the Kepler sequence. We use the inner-edge orbital frequency to identify configurations compatible with a kHz QPO: if the observed QPO traces Keplerian motion in the disk, its frequency must satisfy $\nu_{\rm QPO}\leq\nu_{\rm in}$~\citep{Miller_1998,van_Doesburgh-2018}. Along a fixed-spin sequence, $\nu_{\rm in}$ peaks at the transition, so this condition selects a bounded mass interval. On the surface-terminated branch, $\nu_{\rm in}$ depends on the equatorial radius as well as the mass, and the resulting lower mass boundary inherits the EOS dependence of $R_{\rm eq}$. On the external-ISCO branch, $\nu_{\rm in}$ depends primarily on the mass~\citep{Miller_1998,Luk_2018}, so the resulting upper mass boundary is nearly EOS independent.

As a representative case, Fig.~\ref{fig:mass_radius} includes the constraints for the neutron-star low-mass X-ray binary 4U~0614+09. Its burst oscillations indicate a spin near $414$~Hz~\citep{Strohmayer_2008,Chen_2022,LeStum2026}, and its kHz QPOs reach some of the highest frequencies observed in such systems. Following~\citet{van_Doesburgh-2018}, we adopt the $3\sigma$ lower limit of $1267$~Hz on the centroid frequency of its highest-frequency kHz QPO and require $\nu_{\rm in}\geq1267$~Hz, a conservative choice because a higher QPO frequency would only exclude more configurations. Every DD--PC EOS admits stable configurations satisfying this condition, so the QPO alone does not discriminate among the EOSs. The nuclear constraint nevertheless substantially narrows the allowed stellar properties. The lower mass boundary ranges from $0.81$ to $1.23~M_\odot$ across the full family but only from $0.91$ to $0.98~M_\odot$ within the CNSP-4 band, a reduction of about $82\%$ in its spread, whereas the upper boundary remains at $M\simeq1.97~M_\odot$, varying by less than $0.004~M_\odot$ across the full family. In radius, by contrast, CNSP-4 tightens both boundaries, confining the equatorial radius to $12.40$--$12.71$~km at the lower boundary and $11.78$--$11.81$~km at the upper one, which reduces the spread by about $82\%$ and $95\%$, respectively.

Without the spin measurement, the nonrotating limit gives $M\gtrsim0.71~M_\odot$ for the full family and $M\gtrsim0.81~M_\odot$ within CNSP-4 constraint. Adopting $f=414$~Hz raises these bounds to $0.81$ and $0.91~M_\odot$, respectively. The spin information and nuclear constraint therefore provide complementary restrictions, each increasing the minimum allowed mass by approximately $0.1~M_\odot$. Overall, by implementing CNSP-4 constraint for dipole polarizability, an orbital interpretation of the highest kHz QPO requires $0.91\lesssim M/M_\odot\lesssim1.97$ for 4U~0614+09.

\section{Conclusions}
Measurements of the electric dipole polarizability of neutron-rich nuclei constrain the symmetry energy and, within the DD--PC family of EOSs, narrow the range of masses and equatorial radii at which the prograde ISCO reaches the surface of a rotating neutron star, reducing the EOS-dependent spreads by about $85$--$90\%$. Along this transition, the nuclear constraint also confines the inner-edge orbital frequency to a band narrower than $20$~Hz at each spin considered and reduces the spread in binding efficiency. For an accreting star with a measured spin, a kHz QPO interpreted as disk orbital motion then confines the star to a bounded mass interval with tightly constrained radii. A nuclear observable thus informs both the structure of a rotating neutron star and the motion of matter in its inner accretion disk.

Further accurate measurements of the electric dipole response in neutron-rich nuclei could refine the symmetry-energy constraint underlying these predictions~\citep{Koliogiannis_2026}. On the observational side, the planned enhanced X-ray Timing and Polarimetry mission could detect higher-frequency kHz QPOs~\citep{Zhang_2025_eXTP,Li2025}, which would tighten allowed mass intervals. Spectral observations could provide complementary constraints on the inner disk radius through relativistic disk-line modelling~\citep{Bhattacharyya-2011}. Together with a firmer understanding of the QPO mechanism and disk truncation, these data would allow a more direct comparison with the predicted locations and orbital frequencies of the disk inner edge in sources with measured spins.

\FloatBarrier 

\begin{acknowledgements}
This work was supported by the Croatian Science Foundation under the project number HRZZ-UIP-2026-6884. This work is supported by the Croatian Science Foundation under the project Relativistic Nuclear Many-Body Theory in the Multimessenger Observation Era (HRZZ-IP-2022-10-7773).
This work was supported by the project ``Implementation of cutting-edge research and its application as part of the Scientific Center of Excellence for Quantum and Complex Systems, and Representations of Lie Algebras'', Grant No. PK.1.1.10.0004, co-financed by the European Union through the European Regional Development Fund - Competitiveness and Cohesion Programme 2021-2027. This paper was supported by the European Union – NextGenerationEU through the National Recovery and Resilience Plan 2021-2026 -- Institutional grant of University of Zagreb Faculty of Science (Nuclear Astrophysics). This research was supported by the European Union – NextGenerationEU through the National Recovery and Resilience Plan 2021–2026 Institutional grants of University of Zagreb Faculty of Science (PMF-PRESTIGE).
\end{acknowledgements}

\bibliographystyle{aa}
\bibliography{references}

\begin{thebibliography}{55}
\expandafter\ifx\csname natexlab\endcsname\relax\def\natexlab#1{#1}\fi

\bibitem[{Abbott {et~al.}(2017)Abbott, Abbott, Abbott, Acernese, Ackley, Adams, Adams, Addesso, Adhikari, Adya, Affeldt, Afrough, Agarwal, Agathos, Agatsuma, Aggarwal, Aguiar, Aiello, Ain, Ajith, Allen, Allen, Allocca, Altin, Amato, Ananyeva, Anderson, Anderson, Angelova, Antier, Appert, Arai, Araya, Areeda, Arnaud, Arun, Ascenzi, Ashton, Ast, Aston, Astone, Atallah, Aufmuth, Aulbert, AultONeal, Austin, Avila-Alvarez, Babak, Bacon, Bader, Bae, Bailes, Baker, Baldaccini, Ballardin, Ballmer, Banagiri, Barayoga, Barclay, Barish, Barker, Barkett, Barone, Barr, Barsotti, Barsuglia, Barta, Barthelmy, Bartlett, Bartos, Bassiri, Basti, Batch, Bawaj, Bayley, Bazzan, B{\'e}csy, Beer, Bejger, Belahcene, Bell, Berger, Bergmann, Bernuzzi, Bero, Berry, Bersanetti, Bertolini, Betzwieser, Bhagwat, Bhandare, Bilenko, Billingsley, Billman, Birch, Birney, Birnholtz, Biscans, Biscoveanu, Bisht, Bitossi, Biwer, Bizouard, Blackburn, Blackman, Blair, Blair, Blair, Bloemen, Bock, Bode, Boer, Bogaert, Bohe, Bondu, Bonilla, Bonnand,
  Boom, Bork, Boschi, Bose, Bossie, Bouffanais, Bozzi, Bradaschia, Brady, Branchesi, Brau, Briant, Brillet, Brinkmann, Brisson, Brockill, Broida, Brooks, Brown, Brown, Brunett, Buchanan, Buikema, Bulik, Bulten, Buonanno, Buskulic, Buy, Byer, Cabero, Cadonati, Cagnoli, Cahillane, Calder{\'o}n~Bustillo, Callister, Calloni, Camp, Canepa, Canizares, Cannon, Cao, Cao, Capano, Capocasa, Carbognani, Caride, Carney, Carullo, Casanueva~Diaz, Casentini, Caudill, Cavagli{\`a}, Cavalier, Cavalieri, Cella, Cepeda, Cerd{\'a}-Dur{\'a}n, Cerretani, Cesarini, Chamberlin, Chan, Chao, Charlton, Chase, Chassande-Mottin, Chatterjee, Chatziioannou, Cheeseboro, Chen, Chen, Chen, Cheng, Chia, Chincarini, Chiummo, Chmiel, Cho, Cho, Chow, Christensen, Chu, Chua, Chua, Chung, Chung, Ciani, Ciolfi, Cirelli, Cirone, Clara, Clark, Clearwater, Cleva, Cocchieri, Coccia, Cohadon, Cohen, Colla, Collette, Cominsky, Constancio, Conti, Cooper, Corban, Corbitt, Cordero-Carri{\'o}n, Corley, Cornish, Corsi, Cortese, Costa, Coughlin, Coughlin,
  Coulon, Countryman, Couvares, Covas, Cowan, Coward, Cowart, Coyne, Coyne, Creighton, Creighton, Cripe, Crowder, Cullen, Cumming, Cunningham, Cuoco, Dal~Canton, D{\'a}lya, Danilishin, D'Antonio, Danzmann, Dasgupta, Da~Silva~Costa, Dattilo, Dave, Davier, Davis, Daw, Day, De, DeBra, Degallaix, De~Laurentis, Del{\'e}glise, Del~Pozzo, Demos, Denker, Dent, De~Pietri, Dergachev, De~Rosa, DeRosa, De~Rossi, DeSalvo, de~Varona, Devenson, Dhurandhar, D{\'{\i}}az, Dietrich, Di~Fiore, Di~Giovanni, Di~Girolamo, Di~Lieto, Di~Pace, Di~Palma, Di~Renzo, Doctor, Dolique, Donovan, Dooley, Doravari, Dorrington, Douglas, Dovale~{\'A}lvarez, Downes, Drago, Dreissigacker, Driggers, Du, Ducrot, Dudi, Dupej, Dwyer, Edo, Edwards, Effler, Eggenstein, Ehrens, Eichholz, Eikenberry, Eisenstein, Essick, Estevez, Etienne, Etzel, Evans, Evans, Factourovich, Fafone, Fair, Fairhurst, Fan, Farinon, Farr, Farr, Fauchon-Jones, Favata, Fays, Fee, Fehrmann, Feicht, Fejer, Fernandez-Galiana, Ferrante, Ferreira, Ferrini, Fidecaro, Finstad, Fiori,
  Fiorucci, Fishbach, Fisher, Fitz-Axen, Flaminio, Fletcher, Fong, Font, Forsyth, Forsyth, Fournier, Frasca, Frasconi, Frei, Freise, Frey, Frey, Fries, Fritschel, Frolov, Fulda, Fyffe, Gabbard, Gadre, Gaebel, Gair, Gammaitoni, Ganija, Gaonkar, Garcia-Quiros, Garufi, Gateley, Gaudio, Gaur, Gayathri, Gehrels, Gemme, Genin, Gennai, George, George, Gergely, Germain, Ghonge, Ghosh, Ghosh, Ghosh, Giaime, Giardina, Giazotto, Gill, Glover, Goetz, Goetz, Gomes, Goncharov, Gonz{\'a}lez, Gonzalez~Castro, Gopakumar, Gorodetsky, Gossan, Gosselin, Gouaty, Grado, Graef, Granata, Grant, Gras, Gray, Greco, Green, Gretarsson, Groot, Grote, Grunewald, Gruning, Guidi, Guo, Gupta, Gupta, Gushwa, Gustafson, Gustafson, Halim, Hall, Hall, Hamilton, Hammond, Haney, Hanke, Hanks, Hanna, Hannam, Hannuksela, Hanson, Hardwick, Harms, Harry, Harry, Hart, Haster, Haughian, Healy, Heidmann, Heintze, Heitmann, Hello, Hemming, Hendry, Heng, Hennig, Heptonstall, Heurs, Hild, Hinderer, Ho, Hoak, Hofman, Holt, Holz, Hopkins, Horst, Hough,
  Houston, Howell, Hreibi, Hu, Huerta, Huet, Hughey, Husa, Huttner, Huynh-Dinh, Indik, Inta, Intini, Isa, Isac, Isi, Iyer, Izumi, Jacqmin, Jani, Jaranowski, Jawahar, Jim{\'e}nez-Forteza, Johnson, Johnson-McDaniel, Jones, Jones, Jonker, Ju, Junker, Kalaghatgi, Kalogera, Kamai, Kandhasamy, Kang, Kanner, Kapadia, Karki, Karvinen, Kasprzack, Kastaun, Katolik, Katsavounidis, Katzman, Kaufer, Kawabe, K{\'e}f{\'e}lian, Keitel, Kemball, Kennedy, Kent, Key, Khalili, Khan, Khan, Khan, Khazanov, Kijbunchoo, Kim, Kim, Kim, Kim, Kim, Kim, Kimbrell, King, King, Kinley-Hanlon, Kirchhoff, Kissel, Kleybolte, Klimenko, Knowles, Koch, Koehlenbeck, Koley, Kondrashov, Kontos, Korobko, Korth, Kowalska, Kozak, Kr{\"a}mer, Kringel, Krishnan, Kr{\'o}lak, Kuehn, Kumar, Kumar, Kumar, Kuo, Kutynia, Kwang, Lackey, Lai, Landry, Lang, Lange, Lantz, Lanza, Larson, Lartaux-Vollard, Lasky, Laxen, Lazzarini, Lazzaro, Leaci, Leavey, Lee, Lee, Lee, Lee, Lee, Lehmann, Lenon, Leon, Leonardi, Leroy, Letendre, Levin, Li, Linker, Littenberg, Liu,
  Liu, Lo, Lockerbie, London, Lord, Lorenzini, Loriette, Lormand, Losurdo, Lough, Lousto, Lovelace, L{\"u}ck, Lumaca, Lundgren, Lynch, Ma, Macas, Macfoy, Machenschalk, MacInnis, Macleod, Maga{\~n}a~Hernandez, Maga{\~n}a-Sandoval, Maga{\~n}a~Zertuche, Magee, Majorana, Maksimovic, Man, Mandic, Mangano, Mansell, Manske, Mantovani, Marchesoni, Marion, M{\'a}rka, M{\'a}rka, Markakis, Markosyan, Markowitz, Maros, Marquina, Marsh, Martelli, Martellini, Martin, Martin, Martynov, Marx, Mason, Massera, Masserot, Massinger, Masso-Reid, Mastrogiovanni, Matas, Matichard, Matone, Mavalvala, Mazumder, McCarthy, McClelland, McCormick, McCuller, McGuire, McIntyre, McIver, McManus, McNeill, McRae, McWilliams, Meacher, Meadors, Mehmet, Meidam, Mejuto-Villa, Melatos, Mendell, Mercer, Merilh, Merzougui, Meshkov, Messenger, Messick, Metzdorff, Meyers, Miao, Michel, Middleton, Mikhailov, Milano, Miller, Miller, Miller, Millhouse, Milovich-Goff, Minazzoli, Minenkov, Ming, Mishra, Mitra, Mitrofanov, Mitselmakher, Mittleman, Moffa,
  Moggi, Mogushi, Mohan, Mohapatra, Molina, Montani, Moore, Moraru, Moreno, Morisaki, Morriss, Mours, Mow-Lowry, Mueller, Muir, Mukherjee, Mukherjee, Mukherjee, Mukund, Mullavey, Munch, Mu{\~n}iz, Muratore, Murray, Nagar, Napier, Nardecchia, Naticchioni, Nayak, Neilson, Nelemans, Nelson, Nery, Neunzert, Nevin, Newport, Newton, Ng, Nguyen, Nguyen, Nichols, Nielsen, Nissanke, Nitz, Noack, Nocera, Nolting, North, Nuttall, Oberling, O'Dea, Ogin, Oh, Oh, Ohme, Okada, Oliver, Oppermann, Oram, O'Reilly, Ormiston, Ortega, O'Shaughnessy, Ossokine, Ottaway, Overmier, Owen, Pace, Page, Page, Pai, Pai, Palamos, Palashov, Palomba, Pal-Singh, Pan, Pan, Pang, Pang, Pankow, Pannarale, Pant, Paoletti, Paoli, Papa, Parida, Parker, Pascucci, Pasqualetti, Passaquieti, Passuello, Patil, Patricelli, Pearlstone, Pedraza, Pedurand, Pekowsky, Pele, Penn, Perez, Perreca, Perri, Pfeiffer, Phelps, Piccinni, Pichot, Piergiovanni, Pierro, Pillant, Pinard, Pinto, Pirello, Pitkin, Poe, Poggiani, Popolizio, Porter, Post, Powell, Prasad,
  Pratt, Pratten, Predoi, Prestegard, Prijatelj, Principe, Privitera, Prix, Prodi, Prokhorov, Puncken, Punturo, Puppo, P{\"u}rrer, Qi, Quetschke, Quintero, Quitzow-James, Raab, Rabeling, Radkins, Raffai, Raja, Rajan, Rajbhandari, Rakhmanov, Ramirez, Ramos-Buades, Rapagnani, Raymond, Razzano, Read, Regimbau, Rei, Reid, Reitze, Ren, Reyes, Ricci, Ricker, Rieger, Riles, Rizzo, Robertson, Robie, Robinet, Rocchi, Rolland, Rollins, Roma, Romano, Romano, Romel, Romie, Rosi{\ifmmode \acute{n}\else \'{n}\fi{}}ska, Ross, Rowan, R{\"u}diger, Ruggi, Rutins, Ryan, Sachdev, Sadecki, Sadeghian, Sakellariadou, Salconi, Saleem, Salemi, Samajdar, Sammut, Sampson, Sanchez, Sanchez, Sanchis-Gual, Sandberg, Sanders, Sassolas, Sathyaprakash, Saulson, Sauter, Savage, Sawadsky, Schale, Scheel, Scheuer, Schmidt, Schmidt, Schnabel, Schofield, Sch{\"o}nbeck, Schreiber, Schuette, Schulte, Schutz, Schwalbe, Scott, Scott, Seidel, Sellers, Sengupta, Sentenac, Sequino, Sergeev, Shaddock, Shaffer, Shah, Shahriar, Shaner, Shao, Shapiro,
  Shawhan, Sheperd, Shoemaker, Shoemaker, Siellez, Siemens, Sieniawska, Sigg, Silva, Singer, Singh, Singhal, Sintes, Slagmolen, Smith, Smith, Smith, Somala, Son, Sonnenberg, Sorazu, Sorrentino, Souradeep, Spencer, Srivastava, Staats, Staley, Steinke, Steinlechner, Steinlechner, Steinmeyer, Stevenson, Stone, Stops, Strain, Stratta, Strigin, Strunk, Sturani, Stuver, Summerscales, Sun, Sunil, Suresh, Sutton, Swinkels, Szczepa{\ifmmode \acute{n}\else \'{n}\fi{}}czyk, Tacca, Tait, Talbot, Talukder, Tanner, T{\'a}pai, Taracchini, Tasson, Taylor, Taylor, Tewari, Theeg, Thies, Thomas, Thomas, Thomas, Thorne, Thorne, Thrane, Tiwari, Tiwari, Tokmakov, Toland, Tonelli, Tornasi, Torres-Forn{\'e}, Torrie, T{\"o}yr{\"a}, Travasso, Traylor, Trinastic, Tringali, Trozzo, Tsang, Tse, Tso, Tsukada, Tsuna, Tuyenbayev, Ueno, Ugolini, Unnikrishnan, Urban, Usman, Vahlbruch, Vajente, Valdes, Vallisneri, van Bakel, van Beuzekom, van~den Brand, Van Den~Broeck, Vander-Hyde, van~der Schaaf, van Heijningen, van Veggel, Vardaro, Varma,
  Vass, Vas{\'u}th, Vecchio, Vedovato, Veitch, Veitch, Venkateswara, Venugopalan, Verkindt, Vetrano, Vicer{\'e}, Viets, Vinciguerra, Vine, Vinet, Vitale, Vo, Vocca, Vorvick, Vyatchanin, Wade, Wade, Wade, Walet, Walker, Wallace, Walsh, Wang, Wang, Wang, Wang, Wang, Ward, Warner, Was, Watchi, Weaver, Wei, Weinert, Weinstein, Weiss, Wen, Wessel, We{\ss{}}els, Westerweck, Westphal, Wette, Whelan, Whitcomb, Whiting, Whittle, Wilken, Williams, Williams, Williamson, Willis, Willke, Wimmer, Winkler, Wipf, Wittel, Woan, Woehler, Wofford, Wong, Worden, Wright, Wu, Wysocki, Xiao, Yamamoto, Yancey, Yang, Yap, Yazback, Yu, Yu, Yvert, Zadro{\ifmmode \dot{z}\else \.{z}\fi{}}ny, Zanolin, Zelenova, Zendri, Zevin, Zhang, Zhang, Zhang, Zhang, Zhao, Zhou, Zhou, Zhu, Zhu, Zimmerman, Zucker, \& Zweizig}]{PhysRevLett.119.161101}
Abbott, B.~P., Abbott, R., Abbott, T.~D., {et~al.} 2017, Phys. Rev. Lett., 119, 161101

\bibitem[{Abbott {et~al.}(2019)Abbott, Abbott, Abbott, Acernese, Ackley, Adams, Adams, Addesso, Adhikari, Adya, Affeldt, Agarwal, Agathos, Agatsuma, Aggarwal, Aguiar, Aiello, Ain, Ajith, Allen, Allen, Allocca, Aloy, Altin, Amato, Ananyeva, Anderson, Anderson, Angelova, Antier, Appert, Arai, Araya, Areeda, Ar{\`e}ne, Arnaud, Arun, Ascenzi, Ashton, Ast, Aston, Astone, Atallah, Aubin, Aufmuth, Aulbert, AultONeal, Austin, Avila-Alvarez, Babak, Bacon, Badaracco, Bader, Bae, Baker, Baldaccini, Ballardin, Ballmer, Banagiri, Barayoga, Barclay, Barish, Barker, Barkett, Barnum, Barone, Barr, Barsotti, Barsuglia, Barta, Bartlett, Bartos, Bassiri, Basti, Batch, Bawaj, Bayley, Bazzan, B{\'e}csy, Beer, Bejger, Belahcene, Bell, Beniwal, Bensch, Berger, Bergmann, Bernuzzi, Bero, Berry, Bersanetti, Bertolini, Betzwieser, Bhandare, Bilenko, Bilgili, Billingsley, Billman, Birch, Birney, Birnholtz, Biscans, Biscoveanu, Bisht, Bitossi, Bizouard, Blackburn, Blackman, Blair, Blair, Blair, Bloemen, Bock, Bode, Boer, Boetzel,
  Bogaert, Bohe, Bondu, Bonilla, Bonnand, Booker, Boom, Booth, Bork, Boschi, Bose, Bossie, Bossilkov, Bosveld, Bouffanais, Bozzi, Bradaschia, Brady, Bramley, Branchesi, Brau, Briant, Brighenti, Brillet, Brinkmann, Brisson, Brockill, Brooks, Brown, Brunett, Buchanan, Buikema, Bulik, Bulten, Buonanno, Buskulic, Buy, Byer, Cabero, Cadonati, Cagnoli, Cahillane, Bustillo, Callister, Calloni, Camp, Canepa, Canizares, Cannon, Cao, Cao, Capano, Capocasa, Carbognani, Caride, Carney, Carullo, Diaz, Casentini, Caudill, Cavagli{\`a}, Cavalier, Cavalieri, Cella, Cepeda, Cerd{\'a}-Dur{\'a}n, Cerretani, Cesarini, Chaibi, Chamberlin, Chan, Chao, Charlton, Chase, Chassande-Mottin, Chatterjee, Chatziioannou, Cheeseboro, Chen, Chen, Chen, Cheng, Chia, Chincarini, Chiummo, Chmiel, Cho, Cho, Chow, Christensen, Chu, Chua, Chua, Chung, Chung, Ciani, Ciobanu, Ciolfi, Cipriano, Cirelli, Cirone, Clara, Clark, Clearwater, Cleva, Cocchieri, Coccia, Cohadon, Cohen, Colla, Collette, Collins, Cominsky, Constancio, Conti, Cooper, Corban,
  Corbitt, Cordero-Carri{\'o}n, Corley, Cornish, Corsi, Cortese, Costa, Cotesta, Coughlin, Coughlin, Coulon, Countryman, Couvares, Covas, Cowan, Coward, Cowart, Coyne, Coyne, Creighton, Creighton, Cripe, Crowder, Cullen, Cumming, Cunningham, Cuoco, Canton, D{\'a}lya, Danilishin, D'Antonio, Danzmann, Dasgupta, Costa, Dattilo, Dave, Davier, Davis, Daw, Day, DeBra, Deenadayalan, Degallaix, De~Laurentis, Del{\'e}glise, Del~Pozzo, Demos, Denker, Dent, De~Pietri, Derby, Dergachev, De~Rosa, De~Rossi, DeSalvo, de~Varona, Dhurandhar, D{\'{\i}}az, Dietrich, Di~Fiore, Di~Giovanni, Di~Girolamo, Di~Lieto, Ding, Di~Pace, Di~Palma, Di~Renzo, Dmitriev, Doctor, Dolique, Donovan, Dooley, Doravari, Dorrington, {\'A}lvarez, Downes, Drago, Dreissigacker, Driggers, Du, Dudi, Dupej, Dwyer, Easter, Edo, Edwards, Effler, Eggenstein, Ehrens, Eichholz, Eikenberry, Eisenmann, Eisenstein, Essick, Estelles, Estevez, Etienne, Etzel, Evans, Evans, Fafone, Fair, Fairhurst, Fan, Farinon, Farr, Farr, Fauchon-Jones, Favata, Fays, Fee, Fehrmann,
  Feicht, Fejer, Feng, Fernandez-Galiana, Ferrante, Ferreira, Ferrini, Fidecaro, Fiori, Fiorucci, Fishbach, Fisher, Fishner, Fitz-Axen, Flaminio, Fletcher, Fong, Font, Forsyth, Forsyth, Fournier, Frasca, Frasconi, Frei, Freise, Frey, Frey, Fritschel, Frolov, Fulda, Fyffe, Gabbard, Gadre, Gaebel, Gair, Gammaitoni, Ganija, Gaonkar, Garcia, Garc{\'{\i}}a-Quir{\'o}s, Garufi, Gateley, Gaudio, Gaur, Gayathri, Gemme, Genin, Gennai, George, George, Gergely, Germain, Ghonge, Ghosh, Ghosh, Ghosh, Giacomazzo, Giaime, Giardina, Giazotto, Gill, Giordano, Glover, Goetz, Goetz, Goncharov, Gonz{\'a}lez, Castro, Gopakumar, Gorodetsky, Gossan, Gosselin, Gouaty, Grado, Graef, Granata, Grant, Gras, Gray, Greco, Green, Green, Gretarsson, Groot, Grote, Grunewald, Gruning, Guidi, Gulati, Guo, Gupta, Gupta, Gushwa, Gustafson, Gustafson, Halim, Hall, Hall, Hamilton, Hamilton, Hammond, Haney, Hanke, Hanks, Hanna, Hannam, Hannuksela, Hanson, Hardwick, Harms, Harry, Harry, Hart, Haster, Haughian, Healy, Heidmann, Heintze, Heitmann,
  Hello, Hemming, Hendry, Heng, Hennig, Heptonstall, Hernandez, Heurs, Hild, Hinderer, Hoak, Hochheim, Hofman, Holland, Holt, Holz, Hopkins, Horst, Hough, Houston, Howell, Hreibi, Huerta, Huet, Hughey, Hulko, Husa, Huttner, Huynh-Dinh, Iess, Indik, Ingram, Inta, Intini, Isa, Isac, Isi, Iyer, Izumi, Jacqmin, Jani, Jaranowski, Johnson, Johnson, Jones, Jones, Jonker, Ju, Junker, Kalaghatgi, Kalogera, Kamai, Kandhasamy, Kang, Kanner, Kapadia, Karki, Karvinen, Kasprzack, Kastaun, Katolik, Katsanevas, Katsavounidis, Katzman, Kaufer, Kawabe, Keerthana, K{\'e}f{\'e}lian, Keitel, Kemball, Kennedy, Key, Khalili, Khamesra, Khan, Khan, Khan, Khan, Khazanov, Kijbunchoo, Kim, Kim, Kim, Kim, Kim, Kim, King, King, Kinley-Hanlon, Kirchhoff, Kissel, Kleybolte, Klimenko, Knowles, Koch, Koehlenbeck, Koley, Kondrashov, Kontos, Korobko, Korth, Kowalska, Kozak, Kr{\"a}mer, Kringel, Krishnan, Kr{\'o}lak, Kuehn, Kumar, Kumar, Kumar, Kuo, Kutynia, Kwang, Lackey, Lai, Landry, Landry, Lang, Lange, Lantz, Lanza, Lartaux-Vollard, Lasky,
  Laxen, Lazzarini, Lazzaro, Leaci, Leavey, Lee, Lee, Lee, Lee, Lee, Lehmann, Lenon, Leonardi, Leroy, Letendre, Levin, Li, Li, Li, Linker, Littenberg, Liu, Liu, Lo, Lockerbie, London, Longo, Lorenzini, Loriette, Lormand, Losurdo, Lough, Lousto, Lovelace, L{\"u}ck, Lumaca, Lundgren, Lynch, Ma, Macas, Macfoy, Machenschalk, MacInnis, Macleod, Hernandez, Maga{\~n}a-Sandoval, Zertuche, Magee, Majorana, Maksimovic, Man, Mandic, Mangano, Mansell, Manske, Mantovani, Marchesoni, Marion, M{\'a}rka, M{\'a}rka, Markakis, Markosyan, Markowitz, Maros, Marquina, Martelli, Martellini, Martin, Martin, Martynov, Mason, Massera, Masserot, Massinger, Masso-Reid, Mastrogiovanni, Matas, Matichard, Matone, Mavalvala, Mazumder, McCann, McCarthy, McClelland, McCormick, McCuller, McGuire, McIver, McManus, McRae, McWilliams, Meacher, Meadors, Mehmet, Meidam, Mejuto-Villa, Melatos, Mendell, Mendoza-Gandara, Mercer, Mereni, Merilh, Merzougui, Meshkov, Messenger, Messick, Metzdorff, Meyers, Miao, Michel, Middleton, Mikhailov, Milano,
  Miller, Miller, Miller, Miller, Millhouse, Mills, Milovich-Goff, Minazzoli, Minenkov, Ming, Mishra, Mitra, Mitrofanov, Mitselmakher, Mittleman, Moffa, Mogushi, Mohan, Mohapatra, Montani, Moore, Moraru, Moreno, Morisaki, Mours, Mow-Lowry, Mueller, Muir, Mukherjee, Mukherjee, Mukherjee, Mukund, Mullavey, Munch, Mu{\~n}iz, Muratore, Murray, Nagar, Napier, Nardecchia, Naticchioni, Nayak, Neilson, Nelemans, Nelson, Nery, Neunzert, Nevin, Newport, Ng, Ng, Nguyen, Nguyen, Nichols, Nielsen, Nissanke, Nitz, Nocera, Nolting, North, Nuttall, Obergaulinger, Oberling, O'Brien, O'Dea, Ogin, Oh, Oh, Ohme, Ohta, Okada, Oliver, Oppermann, Oram, O'Reilly, Ormiston, Ortega, O'Shaughnessy, Ossokine, Ottaway, Overmier, Owen, Pace, Pagano, Page, Page, Pai, Pai, Palamos, Palashov, Palomba, Pal-Singh, Pan, Pan, Pang, Pang, Pankow, Pannarale, Pant, Paoletti, Paoli, Papa, Parida, Parker, Pascucci, Pasqualetti, Passaquieti, Passuello, Patil, Patricelli, Pearlstone, Pedersen, Pedraza, Pedurand, Pekowsky, Pele, Penn, Perez, Perreca,
  Perri, Pfeiffer, Phelps, Phukon, Piccinni, Pichot, Piergiovanni, Pierro, Pillant, Pinard, Pinto, Pirello, Pitkin, Poggiani, Popolizio, Porter, Possenti, Post, Powell, Prasad, Pratt, Pratten, Predoi, Prestegard, Principe, Privitera, Prodi, Prokhorov, Puncken, Punturo, Puppo, P{\"u}rrer, Qi, Quetschke, Quintero, Quitzow-James, Raab, Rabeling, Radkins, Raffai, Raja, Rajan, Rajbhandari, Rakhmanov, Ramirez, Ramos-Buades, Rana, Rapagnani, Raymond, Razzano, Read, Regimbau, Rei, Reid, Reitze, Ren, Ricci, Ricker, Riemenschneider, Riles, Rizzo, Robertson, Robie, Robinet, Robson, Rocchi, Rolland, Rollins, Roma, Romano, Romel, Romie, Rosi{\ifmmode \acute{n}\else \'{n}\fi{}}ska, Ross, Rowan, R{\"u}diger, Ruggi, Rutins, Ryan, Sachdev, Sadecki, Sakellariadou, Salconi, Saleem, Salemi, Samajdar, Sammut, Sampson, Sanchez, Sanchez, Sanchis-Gual, Sandberg, Sanders, Sarin, Sassolas, Sathyaprakash, Saulson, Sauter, Savage, Sawadsky, Schale, Scheel, Scheuer, Schmidt, Schnabel, Schofield, Sch{\"o}nbeck, Schreiber, Schuette,
  Schulte, Schutz, Schwalbe, Scott, Scott, Seidel, Sellers, Sengupta, Sentenac, Sequino, Sergeev, Setyawati, Shaddock, Shaffer, Shah, Shahriar, Shaner, Shao, Shapiro, Shawhan, Shen, Shoemaker, Shoemaker, Siellez, Siemens, Sieniawska, Sigg, Silva, Singer, Singh, Singhal, Sintes, Slagmolen, Slaven-Blair, Smith, Smith, Smith, Somala, Son, Sorazu, Sorrentino, Souradeep, Spencer, Srivastava, Staats, Steinke, Steinlechner, Steinlechner, Steinmeyer, Steltner, Stevenson, Stocks, Stone, Stops, Strain, Stratta, Strigin, Strunk, Sturani, Stuver, Summerscales, Sun, Sunil, Suresh, Sutton, Swinkels, Szczepa{\ifmmode \acute{n}\else \'{n}\fi{}}czyk, Tacca, Tait, Talbot, Talukder, Tanner, T{\'a}pai, Taracchini, Tasson, Taylor, Taylor, Tewari, Theeg, Thies, Thomas, Thomas, Thomas, Thorne, Thrane, Tiwari, Tiwari, Tokmakov, Toland, Tonelli, Tornasi, Torres-Forn{\'e}, Torrie, T{\"o}yr{\"a}, Travasso, Traylor, Trinastic, Tringali, Trozzo, Tsang, Tse, Tso, Tsuna, Tsukada, Tuyenbayev, Ueno, Ugolini, Urban, Usman, Vahlbruch, Vajente,
  Valdes, van Bakel, van Beuzekom, van~den Brand, Van Den~Broeck, Vander-Hyde, van~der Schaaf, van Heijningen, van Veggel, Vardaro, Varma, Vass, Vas{\'u}th, Vecchio, Vedovato, Veitch, Veitch, Venkateswara, Venugopalan, Verkindt, Vetrano, Vicer{\'e}, Viets, Vinciguerra, Vine, Vinet, Vitale, Vo, Vocca, Vorvick, Vyatchanin, Wade, Wade, Wade, Walet, Walker, Wallace, Walsh, Wang, Wang, Wang, Wang, Wang, Ward, Warner, Was, Watchi, Weaver, Wei, Weinert, Weinstein, Weiss, Wellmann, Wen, Wessel, We{\ss{}}els, Westerweck, Wette, Whelan, Whiting, Whittle, Wilken, Williams, Williams, Williamson, Willis, Willke, Wimmer, Winkler, Wipf, Wittel, Woan, Woehler, Wofford, Wong, Worden, Wright, Wu, Wysocki, Xiao, Yam, Yamamoto, Yancey, Yang, Yap, Yazback, Yu, Yu, Yvert, Zadro{\ifmmode \dot{z}\else \.{z}\fi{}}ny, Zanolin, Zelenova, Zendri, Zevin, Zhang, Zhang, Zhang, Zhang, Zhang, Zhao, Zhou, Zhou, Zhu, Zhu, Zimmerman, Zlochower, Zucker, \& Zweizig}]{PhysRevX.9.011001}
Abbott, B.~P., Abbott, R., Abbott, T.~D., {et~al.} 2019, Phys. Rev. X, 9, 011001

\bibitem[{Antoniadis {et~al.}(2013)Antoniadis, Freire, Wex, Tauris, Lynch, van Kerkwijk, Kramer, Bassa, Dhillon, Driebe, Hessels, Kaspi, Kondratiev, Langer, Marsh, McLaughlin, Pennucci, Ransom, Stairs, van Leeuwen, Verbiest, \& Whelan}]{doi:10.1126/science.1233232}
Antoniadis, J., Freire, P. C.~C., Wex, N., {et~al.} 2013, Sci., 340, 1233232

\bibitem[{Berti \& Stergioulas(2004)}]{Berti-2004}
Berti, E. \& Stergioulas, N. 2004, MNRAS, 350, 1416

\bibitem[{Bhattacharyya(2011)}]{Bhattacharyya-2011}
Bhattacharyya, S. 2011, MNRAS, 415, 3247

\bibitem[{Birkhan {et~al.}(2017)Birkhan, Miorelli, Bacca, Bassauer, Bertulani, Hagen, Matsubara, von Neumann-Cosel, Papenbrock, Pietralla, Ponomarev, Richter, Schwenk, \& Tamii}]{PhysRevLett.118.252501}
Birkhan, J., Miorelli, M., Bacca, S., {et~al.} 2017, Phys. Rev. Lett., 118, 252501

\bibitem[{Chatziioannou {et~al.}(2025)Chatziioannou, Cromartie, Gandolfi, Tews, Radice, Steiner, \& Watts}]{ymsq-cfcw}
Chatziioannou, K., Cromartie, H.~T., Gandolfi, S., {et~al.} 2025, Rev. Mod. Phys., 97, 045007

\bibitem[{Chen {et~al.}(2022)Chen, Li, Xiong, Ji, Zhang, Peng, Qiao, Li, Wen, Song, Zheng, Song, Zhao, Huang, Lu, Zhang, Xiao, Cai, An, Chang, Chen, Chen, Chen, Dai, Du, Gao, Gong, Guo, Guo, He, Li, Li, Li, Li, Li, Li, Li, Li, Li, Liang, Liang, Liao, Liu, Liu, Liu, Luo, Ma, Meng, Ou, Shi, Shi, Sun, Sun, Tuo, Wang, Wang, Wang, Wang, Wang, Wang, Wang, Wang, Wen, Wu, Xie, Xu, Xu, Xue, Yang, Yao, Ye, Yi, Zhang, Zhang, Zhang, Zhang, Zhang, Zhang, Zhang, Zhang, Zhang, Zhang, Zhang, Zhao, Zhao, Zhao, Zheng, Zhou, \& Zhu}]{Chen_2022}
Chen, Y.-P., Li, J., Xiong, S.-L., {et~al.} 2022, ApJ, 935, 10

\bibitem[{Choudhury {et~al.}(2024)Choudhury, Salmi, Vinciguerra, Riley, Kini, Watts, Dorsman, Bogdanov, Guillot, Ray, Reardon, Remillard, Bilous, Huppenkothen, Lattimer, Rutherford, Arzoumanian, Gendreau, Morsink, \& Ho}]{Choudhury_2024}
Choudhury, D., Salmi, T., Vinciguerra, S., {et~al.} 2024, \apjl, 971, L20

\bibitem[{Cipolletta {et~al.}(2017)Cipolletta, Cherubini, Filippi, Rueda, \& Ruffini}]{PhysRevD.96.024046}
Cipolletta, F., Cherubini, C., Filippi, S., Rueda, J.~A., \& Ruffini, R. 2017, Phys. Rev. D, 96, 024046

\bibitem[{{Cook} {et~al.}(1994){Cook}, {Shapiro}, \& {Teukolsky}}]{Cook-1994}
{Cook}, G.~B., {Shapiro}, S.~L., \& {Teukolsky}, S.~A. 1994, ApJ, 424, 823

\bibitem[{Fonseca {et~al.}(2021)Fonseca, Cromartie, Pennucci, Ray, Kirichenko, Ransom, Demorest, Stairs, Arzoumanian, Guillemot, Parthasarathy, Kerr, Cognard, Baker, Blumer, Brook, DeCesar, Dolch, Dong, Ferrara, Fiore, Garver-Daniels, Good, Jennings, Jones, Kaspi, Lam, Lorimer, Luo, McEwen, McKee, McLaughlin, McMann, Meyers, Naidu, Ng, Nice, Pol, Radovan, Shapiro-Albert, Tan, Tendulkar, Swiggum, Wahl, \& Zhu}]{Fonseca_2021}
Fonseca, E., Cromartie, H.~T., Pennucci, T.~T., {et~al.} 2021, \apjl, 915, L12

\bibitem[{{Friedman} {et~al.}(1988){Friedman}, {Ipser}, \& {Sorkin}}]{Friedman-1988}
{Friedman}, J.~L., {Ipser}, J.~R., \& {Sorkin}, R.~D. 1988, ApJ, 325, 722

\bibitem[{Hashimoto {et~al.}(2015)Hashimoto, Krumbholz, Reinhard, Tamii, von Neumann-Cosel, Adachi, Aoi, Bertulani, Fujita, Fujita, Ganio{\ifmmode \check{g}\else \v{g}\fi{}}lu, Hatanaka, Ideguchi, Iwamoto, Kawabata, Khai, Krugmann, Martin, Matsubara, Miki, Neveling, Okamura, Ong, Poltoratska, Ponomarev, Richter, Sakaguchi, Shimbara, Shimizu, Simonis, Smit, S{\"u}soy, Suzuki, Thies, Yosoi, \& Zenihiro}]{PhysRevC.92.031305}
Hashimoto, T., Krumbholz, A.~M., Reinhard, P.-G., {et~al.} 2015, Phys. Rev. C, 92, 031305(R)

\bibitem[{Hessels {et~al.}(2006)Hessels, Ransom, Stairs, Freire, Kaspi, \& Camilo}]{doi:10.1126/science.1123430}
Hessels, J. W.~T., Ransom, S.~M., Stairs, I.~H., {et~al.} 2006, Sci., 311, 1901

\bibitem[{Kaaret {et~al.}(1997)Kaaret, Ford, \& Chen}]{Kaaret_1997}
Kaaret, P., Ford, E.~C., \& Chen, K. 1997, ApJ, 480, L27

\bibitem[{Klis(2000)}]{Klis_2000}
Klis, M. v.~d. 2000, Annu. Rev. Astron. Astrophys., 38, 717

\bibitem[{{Kluzniak} \& {Wagoner}(1985)}]{Kluzniak-1985}
{Kluzniak}, W. \& {Wagoner}, R.~V. 1985, ApJ, 297, 548

\bibitem[{Kluźniak(1998)}]{Kluzniak_1998}
Kluźniak, W. 1998, ApJ, 509, L37

\bibitem[{Koliogiannis {et~al.}(2026{\natexlab{a}})Koliogiannis, Yüksel, Ghosh, \& Paar}]{Koliogiannis_2026}
Koliogiannis, P., Yüksel, E., Ghosh, T., \& Paar, N. 2026{\natexlab{a}}, \apjl, 996, L18

\bibitem[{Koliogiannis {et~al.}(2025)Koliogiannis, Yüksel, \& Paar}]{KOLIOGIANNIS2025139362}
Koliogiannis, P., Yüksel, E., \& Paar, N. 2025, Phys. Lett. B, 862, 139362

\bibitem[{Koliogiannis {et~al.}(2026{\natexlab{b}})Koliogiannis, Ghosh, Y\"uksel, \& Paar}]{9mnb-t3yz}
Koliogiannis, P.~S., Ghosh, T., Y\"uksel, E., \& Paar, N. 2026{\natexlab{b}}, Phys. Rev. C, 113, 055809

\bibitem[{Lattimer \& Prakash(2001)}]{Lattimer_2001}
Lattimer, J.~M. \& Prakash, M. 2001, ApJ, 550, 426

\bibitem[{Le~Stum {et~al.}(2026)Le~Stum, Cangemi, Coleiro, Guillot, Chenevez, Bacon, Bellemont, Bouchet, Bouchet, Cavet, Cordier, Foisseau, Godet, Goldwurm, Han, Lachaud, Li, Li, Qiu, Rodriguez, Tan, Tao, Verwaerde, Wang, Wang, Wei, Wu, Xie, Xin, Xiong, Zhang, \& Zheng}]{LeStum2026}
Le~Stum, S., Cangemi, F., Coleiro, A., {et~al.} 2026, \apjl, 997, L25

\bibitem[{Li {et~al.}(2025)Li, Watts, Zhang, {et~al.}}]{Li2025}
Li, A., Watts, A.~L., Zhang, G., {et~al.} 2025, Sci. China Phys. Mech. Astron., 68, 119503

\bibitem[{Luk \& Lin(2018)}]{Luk_2018}
Luk, S.-S. \& Lin, L.-M. 2018, ApJ, 861, 141

\bibitem[{Miller {et~al.}(1998)Miller, Lamb, \& Cook}]{Miller_1998}
Miller, M.~C., Lamb, F.~K., \& Cook, G.~B. 1998, ApJ, 509, 793

\bibitem[{Miller {et~al.}(2019)Miller, Lamb, Dittmann, Bogdanov, Arzoumanian, Gendreau, Guillot, Harding, Ho, Lattimer, Ludlam, Mahmoodifar, Morsink, Ray, Strohmayer, Wood, Enoto, Foster, Okajima, Prigozhin, \& Soong}]{Miller_2019}
Miller, M.~C., Lamb, F.~K., Dittmann, A.~J., {et~al.} 2019, \apjl, 887, L24

\bibitem[{Miller {et~al.}(2021)Miller, Lamb, Dittmann, Bogdanov, Arzoumanian, Gendreau, Guillot, Ho, Lattimer, Loewenstein, Morsink, Ray, Wolff, Baker, Cazeau, Manthripragada, Markwardt, Okajima, Pollard, Cognard, Cromartie, Fonseca, Guillemot, Kerr, Parthasarathy, Pennucci, Ransom, \& Stairs}]{Miller_2021}
Miller, M.~C., Lamb, F.~K., Dittmann, A.~J., {et~al.} 2021, \apjl, 918, L28

\bibitem[{Nikšić {et~al.}(2014)Nikšić, Paar, Vretenar, \& Ring}]{NIKSIC20141808}
Nikšić, T., Paar, N., Vretenar, D., \& Ring, P. 2014, CoPhC, 185, 1808

\bibitem[{Oertel {et~al.}(2017)Oertel, Hempel, Kl\"ahn, \& Typel}]{Oertel2017}
Oertel, M., Hempel, M., Kl\"ahn, T., \& Typel, S. 2017, Rev. Mod. Phys., 89, 015007

\bibitem[{Pappas(2015)}]{Pappas-2015}
Pappas, G. 2015, MNRAS, 454, 4066

\bibitem[{Paschalidis \& Stergioulas(2017)}]{Paschalidis_2017}
Paschalidis, V. \& Stergioulas, N. 2017, Living Rev. Relativ., 20, 7

\bibitem[{Psaltis(2008)}]{Psaltis-2008}
Psaltis, D. 2008, Living Rev. Relativ., 11, 9

\bibitem[{Riley {et~al.}(2019)Riley, Watts, Bogdanov, Ray, Ludlam, Guillot, Arzoumanian, Baker, Bilous, Chakrabarty, Gendreau, Harding, Ho, Lattimer, Morsink, \& Strohmayer}]{Riley_2019}
Riley, T.~E., Watts, A.~L., Bogdanov, S., {et~al.} 2019, \apjl, 887, L21

\bibitem[{Riley {et~al.}(2021)Riley, Watts, Ray, Bogdanov, Guillot, Morsink, Bilous, Arzoumanian, Choudhury, Deneva, Gendreau, Harding, Ho, Lattimer, Loewenstein, Ludlam, Markwardt, Okajima, Prescod-Weinstein, Remillard, Wolff, Fonseca, Cromartie, Kerr, Pennucci, Parthasarathy, Ransom, Stairs, Guillemot, \& Cognard}]{Riley_2021}
Riley, T.~E., Watts, A.~L., Ray, P.~S., {et~al.} 2021, \apjl, 918, L27

\bibitem[{Roca-Maza \& Paar(2018)}]{ROCAMAZA201896}
Roca-Maza, X. \& Paar, N. 2018, Prog. Part. Nucl. Phys., 101, 96

\bibitem[{Romani {et~al.}(2022)Romani, Kandel, Filippenko, Brink, \& Zheng}]{Romani_2022}
Romani, R.~W., Kandel, D., Filippenko, A.~V., Brink, T.~G., \& Zheng, W. 2022, \apjl, 934, L17

\bibitem[{Rossi {et~al.}(2013)Rossi, Adrich, Aksouh, Alvarez-Pol, Aumann, Benlliure, B{\"o}hmer, Boretzky, Casarejos, Chartier, Chatillon, Cortina-Gil, Datta~Pramanik, Emling, Ershova, Fernandez-Dominguez, Geissel, Gorska, Heil, Johansson, Junghans, Kelic-Heil, Kiselev, Klimkiewicz, Kratz, Kr{\"u}cken, Kurz, Labiche, Le~Bleis, Lemmon, Litvinov, Mahata, Maierbeck, Movsesyan, Nilsson, Nociforo, Palit, Paschalis, Plag, Reifarth, Savran, Scheit, Simon, S{\"u}mmerer, Wagner, Walu{\ifmmode \acute{s}\else \'{s}\fi{}}, Weick, \& Winkler}]{PhysRevLett.111.242503}
Rossi, D.~M., Adrich, P., Aksouh, F., {et~al.} 2013, Phys. Rev. Lett., 111, 242503

\bibitem[{Salmi {et~al.}(2024)Salmi, Deneva, Ray, Watts, Choudhury, Kini, Vinciguerra, Cromartie, Wolff, Arzoumanian, Bogdanov, Gendreau, Guillot, Ho, Morsink, Cognard, Guillemot, Theureau, \& Kerr}]{Salmi_2024}
Salmi, T., Deneva, J.~S., Ray, P.~S., {et~al.} 2024, \apj, 976, 58

\bibitem[{Schaab \& Weigel(1999)}]{Schaab_1999}
Schaab, C. \& Weigel, M.~K. 1999, MNRAS, 308, 718

\bibitem[{Shapiro \& Teukolsky(1983)}]{Shapiro-1983}
Shapiro, S. \& Teukolsky, S. 1983, Black {H}oles, {W}hite {D}warfs, and {N}eutron Stars (New York: John Wiley and Sons)

\bibitem[{Shibata \& Sasaki(1998)}]{PhysRevD.58.104011}
Shibata, M. \& Sasaki, M. 1998, Phys. Rev. D, 58, 104011

\bibitem[{{Sibgatullin} \& {Sunyaev}(1998)}]{Sibgatullin_1998}
{Sibgatullin}, N.~R. \& {Sunyaev}, R.~A. 1998, Astronomy Letters, 24, 774

\bibitem[{{Stergioulas} \& {Friedman}(1995)}]{Stergioulas-1995}
{Stergioulas}, N. \& {Friedman}, J.~L. 1995, ApJ, 444, 306

\bibitem[{Strohmayer {et~al.}(2007)Strohmayer, Markwardt, \& Kuulkers}]{Strohmayer_2008}
Strohmayer, T.~E., Markwardt, C.~B., \& Kuulkers, E. 2007, ApJ, 672, L37

\bibitem[{Tamii {et~al.}(2011)Tamii, Poltoratska, von Neumann-Cosel, Fujita, Adachi, Bertulani, Carter, Dozono, Fujita, Fujita, Hatanaka, Ishikawa, Itoh, Kawabata, Kalmykov, Krumbholz, Litvinova, Matsubara, Nakanishi, Neveling, Okamura, Ong, {\"O}zel-Tashenov, Ponomarev, Richter, Rubio, Sakaguchi, Sakemi, Sasamoto, Shimbara, Shimizu, Smit, Suzuki, Tameshige, Wambach, Yamada, Yosoi, \& Zenihiro}]{PhysRevLett.107.062502}
Tamii, A., Poltoratska, I., von Neumann-Cosel, P., {et~al.} 2011, Phys. Rev. Lett., 107, 062502

\bibitem[{Thampan {et~al.}(1999)Thampan, Bhattacharya, \& Datta}]{Thampan-1999}
Thampan, A.~V., Bhattacharya, D., \& Datta, B. 1999, MNRAS, 302, L69

\bibitem[{Thampan \& Datta(1998)}]{Thampan_1998}
Thampan, A.~V. \& Datta, B. 1998, MNRAS, 297, 570

\bibitem[{Török {et~al.}(2014)Török, Urbanec, Adámek, \& Urbancová}]{Torok-2014}
Török, G., Urbanec, M., Adámek, K., \& Urbancová, G. 2014, A\&A, 564, L5

\bibitem[{van Doesburgh {et~al.}(2018)van Doesburgh, van der Klis, \& Morsink}]{van_Doesburgh-2018}
van Doesburgh, M., van der Klis, M., \& Morsink, S.~M. 2018, MNRAS, 479, 426

\bibitem[{Yunes {et~al.}(2022)Yunes, Miller, \& Yagi}]{Yunes_2022}
Yunes, N., Miller, M.~C., \& Yagi, K. 2022, Nat. Rev. Phys., 4, 237

\bibitem[{Yüksel {et~al.}(2021)Yüksel, Oishi, \& Paar}]{Yuksel2021}
Yüksel, E., Oishi, T., \& Paar, N. 2021, Universe, 7, 71

\bibitem[{Zhang {et~al.}(2025)Zhang, Santangelo, Xu, {et~al.}}]{Zhang_2025_eXTP}
Zhang, S.-N., Santangelo, A., Xu, Y., {et~al.} 2025, Sci. China Phys. Mech. Astron., 68, 119502

\bibitem[{Zhang {et~al.}(1997)Zhang, Strohmayer, \& Swank}]{Zhang_1997}
Zhang, W., Strohmayer, T.~E., \& Swank, J.~H. 1997, ApJ, 482, L167

\end{thebibliography}

\end{document}